\documentclass[russian,english,prd,aps,final,notitlepage,nobibnotes,nofootinbib,superscriptaddress,showpacs]{revtex4}
\usepackage[T1]{fontenc}
\usepackage[koi8-r]{inputenc}
\usepackage{array}
\usepackage{multirow}
\usepackage{graphicx}
\usepackage{esint}

\makeatletter

\providecommand{\tabularnewline}{\\}

\@ifundefined{textcolor}{}
{%
 \definecolor{BLACK}{gray}{0}
 \definecolor{WHITE}{gray}{1}
 \definecolor{RED}{rgb}{1,0,0}
 \definecolor{GREEN}{rgb}{0,1,0}
 \definecolor{BLUE}{rgb}{0,0,1}
 \definecolor{CYAN}{cmyk}{1,0,0,0}
 \definecolor{MAGENTA}{cmyk}{0,1,0,0}
 \definecolor{YELLOW}{cmyk}{0,0,1,0}
}

\newcommand{\GeV}{\mathrm{GeV}}
 
\newcommand{\Br}{\mathrm{Br}}

\newcommand{\R}{\mathcal{R}}

\newcommand{\M}{\mathcal{M}}
\newcommand{\eW}{\varepsilon^{(W)}}
\newcommand{\ch}{\mathrm{ch}}

\newcommand{\brJpppSR}{0.39}

\newcommand{\brJpppppSR}{0.18}
\newcommand{\brJpiSR}{0.17}

\newcommand{\brJpppPM}{0.45}

\newcommand{\brJpppppPM}{0.2}
\newcommand{\brJpiPM}{0.21}

\newcommand{\brJpppPMM}{0.14}

\newcommand{\brJpppppPMM}{0.066}
\newcommand{\brJpiPMM}{0.064}

\newcommand{\brJJpppppSR}{3.5\times 10^{-3}}
\newcommand{\brJJpiSR}{6.6\times 10^{-3}}

\newcommand{\brJJpppppPM}{3.9\times 10^{-3}}
\newcommand{\brJJpiPM}{7.8\times 10^{-3}}

\newcommand{\brJJpppppPMM}{6.9\times 10^{-4}}
\newcommand{\brJJpiPMM}{0.014}

\newcommand{\brBSSpppppSR}{1.9\times 10^{-5}}
\newcommand{\brBSSpiSR}{7.7}

\newcommand{\brBSSpppppPM}{2.\times 10^{-5}}
\newcommand{\brBSSpiPM}{9.6}

\newcommand{\brBSpppppSR}{9.1\times 10^{-6}}
\newcommand{\brBSpiSR}{17.}

\newcommand{\brBSpppppPM}{6.5\times 10^{-6}}
\newcommand{\brBSpiPM}{12.}
\newcommand{\ratioJpppSR}{2.3}
\newcommand{\ratioJpppppSR}{1.1}
\newcommand{\ratioJpppPM}{2.2}
\newcommand{\ratioJpppppPM}{0.95}
\newcommand{\ratioJpppPMM}{2.1}
\newcommand{\ratioJpppppPMM}{1.}

\newcommand{\ratioJJpppppSR}{0.53}

\newcommand{\ratioJJpppppPM}{0.5}

\newcommand{\ratioJJpppppPMM}{0.05}

\newcommand{\ratioBSALLpppppSR}{1.1\times 10^{-6}}

\newcommand{\ratioBSALLpppppPM}{1.2\times 10^{-6}}

\usepackage{babel}\addto\captionsenglish{}
\addto\captionsenglish{}

\makeatother

\usepackage{babel}
\begin{document}

\title{Production of charged $\pi$-mesons in exclusive $B_{c}\to V(P)+n\pi$
decays}

\author{A.V. Luchinsky}

\email{Alexey.Luchinsky@ihep.ru}

\affiliation{Institute for High Energy Physics, Protvino, Russia}
\begin{abstract}
The paper is devoted to vector or pseudoscalar heavy quarkonia production
$\psi^{(')}$, $B_{s}^{(*)}$ in association with five charged $\pi$-mesons
in exclusive $B_{c}$-meson decays. Using available formfactors parameterizations
and spectral functions of $W\to5\pi$ transition we obtain branching
fractions of these decays and distributions over invariant mass of
$(5\pi$)-system. 
\end{abstract}

\pacs{13.25.Hw, 14.40.Pq, 12.39.St}

\maketitle

\section{Introduction}

Heavy quarkonia (for example charmonium mesons $\eta_{c}$, $J/\psi$,
$\psi(2S)$, etc, or bottomonia $\eta_{b}$, $\Upsilon$, $\chi_{bJ}$,
\dots) always played a special role in elementary particle physics.
Due to the presence of a heavy quark with mass $m_{Q}\gg\Lambda_{\mathrm{QCD}}$
the scales of quark-antiquark annihilation ($\sim1/m_{Q}$) and hadronization
into experimentally observed meson ($\sim1/\Lambda_{\mathrm{QCD}}$)
differ significantly and these two processes are separated. As a result
the reactions of heavy quarkonia production and decays can be used
for analysis of strong interaction both in perturbative and nonperturbative
regimes. Over the last years significant theoretical and experimental
results were obtained in this field.

Heavy quarkonia with open flavor $\left(b\bar{c}\right)$, i.e. $B_{c}$-meson
and its excitations, take intermediate place between charmonium and
bottomonium mesons. So, they provide a possibility for independent
test of models used in analysis of states with hidden flavor. Theoretical
predictions for the width and lifetime of $B_{c}$-meson\cite{Gershtein:1994jw}
\begin{eqnarray}
M_{B_{c}} & = & 6.25\,\GeV,\qquad\tau_{B_{c}}=0.46\,\mathrm{ps}
\end{eqnarray}
 are in good agreement with experimental values \cite{Aaltonen:2007gv}.
Predictions for partial widths, on the other hand, differ significantly
from experimental results. For example, ratios presented in ref. \cite{Papadimitriou:2005qt}
\begin{eqnarray}
\frac{\sigma_{B_{c}}\Br\left(B_{c}\to J/\psi e^{+}\nu_{e}\right)}{\sigma_{B}\Br\left(B\to J/\psi K\right)} & = & 0.282\pm0.038\pm0.074
\end{eqnarray}
 and 
\begin{eqnarray}
\frac{\sigma_{B_{c}}\Br\left(B_{c}\to J/\psi\mu^{+}\nu_{\mu}\right)}{\sigma_{B}\Br\left(B\to J/\psi K\right)} & = & 0.249\pm0.045_{-0.076}^{+0.107}
\end{eqnarray}
are approximately an order of magnitude higher than theoretical expectations
based on current estimates for $B_{c}$-meson production cross section
and the branching fraction of its semileptonic decay \cite{Gershtein:1994jw}.

In the present paper we consider exclusive decays $B_{c}\to V(P)+\R$,
where $\R$ is a set of light mesons (e.g. $5\pi$). According to
factorization theorem the widths of these decays are connected directly
with the widths of $\tau$-lepton decays $\tau\to\nu_{\tau}+\R$.
In both cases the final state $\R$ is controlled by virtual $W$-boson
that is produced in heavy quark weak decays for hadronic processes
or $\tau\to\nu_{\tau}W$ in leptonic ones. In contrast to $\tau$-lepton
decays, $B_{c}\to V(P)W$ transition is described by formfactors,
so it is possible to determine these formfactors using available experimental
data about $\tau\to\nu_{\tau}\R$ and $B_{c}\to V(P)\R$ decays and
compare them with theoretical predictions based on QCD sum rules and
different potential models. Analysis of $B_{c}\to V(P)\R$ decays
can also be used to describe $W\to\R$ transitions in energy regions
that cannot be achieved in $\tau$-lepton decays.

In the next section of our paper we present analytical expressions
for $B_{c}\to V(P)\R$ widths and distributions over invariant mass
of light meson system $\R$. Technique of spectral functions used
in our paper is also described briefly in that section. In section
\ref{sec:Num} we give numerical expressions for $B_{c}\to V(P)$
fromfactors, $W\to\R$ spectral functions and present our predictions
for branching fractions of decays and $q^{2}$ distributions considered
in our article. Brief discussion of the results is given in the last
section.

\section{Analytical Results}

\begin{figure}
\begin{centering}
\includegraphics[width=0.9\textwidth]{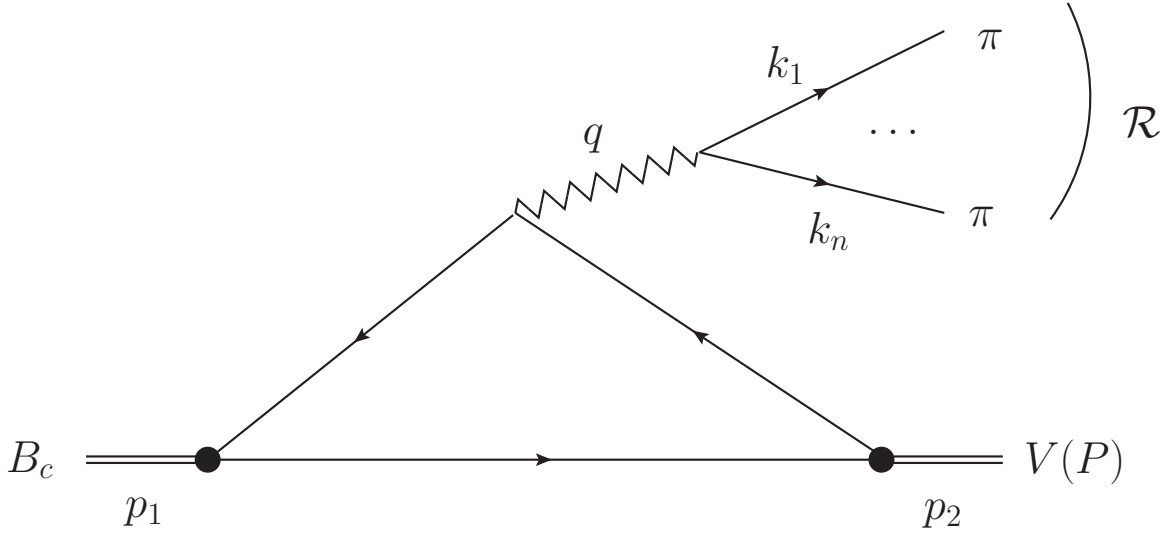} 
\par\end{centering}

\caption{Typical diagram for $B_{c}\to V(P)+n\pi$ process\label{diag}}
\end{figure}

It is well known that in the valence approximation $B_{c}$-mesons
consist of $b$- and $c$-quarks, so their decays into heavy quarkonia
are caused by weak decays of one of the constituent quarks: $b\to cW^{*}$
for charmonium meson in the final state or $c\to sW^{*}$ in the case
of $B_{s}^{(*)}$ production. Produced in this decay heavy quark in
combination with other $B_{c}$ constituent forms final heavy quarkonium,
while virtual $W$-boson hadronizes into light mesons system $\R$.
Typical diagram of this process is shown in fig.\ref{diag}.

In our paper we consider decays

\begin{eqnarray}
B_{c} & \to & V(P)+\R,\label{eq:dec}
\end{eqnarray}
 where $V(P)$ is vector (pseudoscalar) heavy quarkonium $J/\psi$,
$\psi(2S)$ or $B_{s}^{(*)}$, and $\R=(5\pi)_{\ch}$is a set of five
charged $\pi$-mesons (production of smaller number of $\pi$'s was
considered in previous articles \cite{Likhoded:2009ib,Likhoded:2010jr,Wang2012}).
The amplitude of the reaction (\ref{eq:dec}) in the framework of
factorization theorem can be expressed through formfactors of $B_{c}\to V(P)$
transition and is equal to 
\begin{eqnarray}
\M_{V(P)} & = & \frac{G_{F}V_{ij}}{\sqrt{2}}a_{1}H_{V(P)}^{\mu}\eW_{\mu},\label{eq:amp}
\end{eqnarray}
 where $V_{ij}$ is the element of CKM mixing matrix, $\eW$ is the
effective polarization vector of virtual $W$-boson, coefficient $a_{1}$
describes the effect of soft gluon rescattering \cite{Buchalla:1995vs},
and $H_{V(P)}$ is the amplitude of $B_{c}\to V(P)$ transition. This
amplitude can be written in the form

\begin{eqnarray}
H_{V}^{\mu} & = & 2M_{2}A_{0}\left(q^{2}\right)\frac{q^{\mu}\left(q\epsilon\right)}{q^{2}}+\left(M_{1}+M_{2}\right)A_{1}\left(q^{2}\right)\left(\epsilon^{\mu}-\frac{q^{\mu}(q\epsilon)}{q^{2}}\right)-\nonumber \\
 & - & A_{2}\left(q^{2}\right)\frac{(q\epsilon)}{M_{1}+M_{2}}\left(p_{1}^{\mu}+p_{2}^{\mu}-\frac{M_{1}^{2}-M_{2}^{2}}{q^{2}}q^{\mu}\right)-\frac{2iV\left(q^{2}\right)}{M_{1}+M_{2}}e_{\mu\nu\alpha\beta}\epsilon^{\nu}p_{1}^{\alpha}p_{2}^{\beta}\label{eq:AV}
\end{eqnarray}
 for vector meson production and

\begin{eqnarray}
H_{\mu}^{P} & = & F_{+}\left(q^{2}\right)\left(p_{1}^{\mu}+p_{2}^{\mu}-\frac{M_{1}^{2}-M_{2}^{2}}{q^{2}}q^{\mu}\right)+F_{0}\left(q^{2}\right)\frac{M_{1}^{2}-M_{2}^{2}}{q^{2}}q^{\mu}\label{eq:AP}
\end{eqnarray}
 in the case of pseudoscalar meson in the final state. In the above
expressions $p_{1,2}$ and $M_{1,2}$ are momenta and masses of initial
and final heavy quarkonia, $\epsilon^{\mu}$ is the polarization vector
of final vector meson, $q=p_{1}-p_{2}$ is the momentum of virtual
$W$-boson, $F_{0,+}\left(q^{2}\right)$, $A_{0,1,2}\left(q^{2}\right)$
and $V\left(q^{2}\right)$ are formfactors. It is clear that these
formfactors cannot be determined from perturbative QCD, so one should
apply some nonperturbative methods. In our paper we use the results
based on QCD sum rules \cite{Kiselev:2002vz} and solution of different
potential models \cite{Kiselev:2002vz,Kiselev:1993eb,Kiselev:1994ay},
\cite{Ebert:2003cn}. Further these formfactor sets will be denoted
as SR, PM1, and PM2 respectively.

If we are interested only in partial widths of the considered decays
and $q^{2}$-distributions, it is convenient to integrate over the
phase space of light mesons system and use the technique of spectral
functions (see ref.\cite{Schael:2005am} for detailed description).
In the framework of this approach the differential widths of $B_{c}$-meson
decays into vector or pseudoscalar quarkonium have the form

\begin{eqnarray}
\frac{d\Gamma\left[B_{c}\to V\R\right]}{dq^{2}} & = & \frac{G_{F}^{2}V_{cb}^{2}a_{1}^{2}M_{1}^{3}\left|V_{ij}\right|^{2}}{128\pi M_{2}^{2}\left(M_{1}+M_{2}\right)^{2}}\beta^{3}\times\nonumber \\
 &  & \left\{ \rho_{T}^{\R}\left[\left(M_{1}+M_{2}\right)^{2}\left(1+\frac{12M_{2}^{2}q^{2}}{M_{1}^{4}\beta^{2}}\right)A_{1}^{2}+M_{1}^{4}\beta^{2}A_{2}^{2}+8M_{2}^{2}q^{2}V^{2}-\right.\right.\nonumber \\
 &  & \left.\left.2\left(M_{1}+M_{2}\right)^{2}\left(M_{1}^{2}-M_{2}^{2}-q^{2}\right)A_{1}A_{2}\right]+4A_{0}^{2}\left(M_{1}+M_{2}\right)^{2}\rho_{L}^{\R}\right\} ,\label{eq:dGV}
\end{eqnarray}
 and 
\begin{eqnarray}
\frac{d\Gamma\left[B_{c}\to P\R\right]}{dq} & = & \frac{G_{F}^{2}a_{1}^{2}\left|V_{ij}\right|^{2}}{32\pi M_{1}}\beta\left\{ \left(M_{1}^{2}-M_{2}^{2}\right)^{2}F_{0}^{2}\rho_{L}^{\R}+M_{1}^{4}\beta^{2}F_{+}\rho_{T}^{\R}\right\} .\label{eq:dGP}
\end{eqnarray}
 respectively. In these expressions $\beta$ is the velocity of final
quarkonium in $B_{c}$-meson rest frame

\begin{eqnarray}
\beta & = & \sqrt{1-\left(\frac{M_{2}+\sqrt{q^{2}}}{M_{1}}\right)^{2}}\sqrt{1-\left(\frac{M_{2}-\sqrt{q^{2}}}{M_{1}}\right)^{2}},
\end{eqnarray}
 and $\rho_{L,T}^{\R}\left(q^{2}\right)$ are longitudinal and transverse
spectral functions defined according to

\begin{eqnarray}
\left(q_{\mu}q_{\nu}-q^{2}g_{\mu\nu}\right)\rho_{T}\left(q^{2}\right)+q_{\mu}q_{\nu}\rho_{L}\left(q^{2}\right) & = & \frac{1}{2\pi}\int d\Phi_{n}\left(q\to\R\right)\eW_{\mu}(\eW_{\nu})^{*},
\end{eqnarray}
 where 
\begin{eqnarray}
d\Phi_{n}\left(q\to\R\right) & = & \left(2\pi\right)^{4}\delta^{4}\left(q-\sum k_{i}\right)\prod\frac{d^{3}k_{i}}{\left(2\pi\right)^{3}2E_{i}}
\end{eqnarray}
 is the Lorentz-invariant phase space of system $\R$.

\section{Numerical Results\label{sec:Num}}

Let us first consider explicit expressions for spectral functions
of different final states that will be used in our article.

The amplitude of $W\to\pi$ transition has the form

\begin{eqnarray}
\left\langle \pi(k)\left|J_{\mu}\right|W\right\rangle  & = & f_{\pi}k_{\mu},
\end{eqnarray}
 where $f_{\pi}\approx130$ MeV. Resulting transverse and longitudinal
spectral functions are equal to

\begin{eqnarray}
\rho_{T}^{(\pi)}\left(q^{2}\right) & = & 0,\qquad\rho_{L}^{(\pi)}\left(q^{2}\right)=f_{\pi}^{2}.
\end{eqnarray}

\begin{figure}
\begin{centering}
\includegraphics[width=0.4\textwidth]{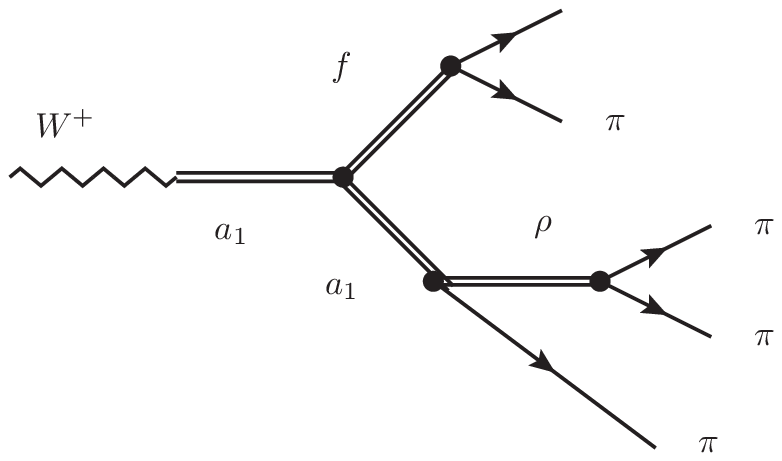}\includegraphics[width=0.4\textwidth]{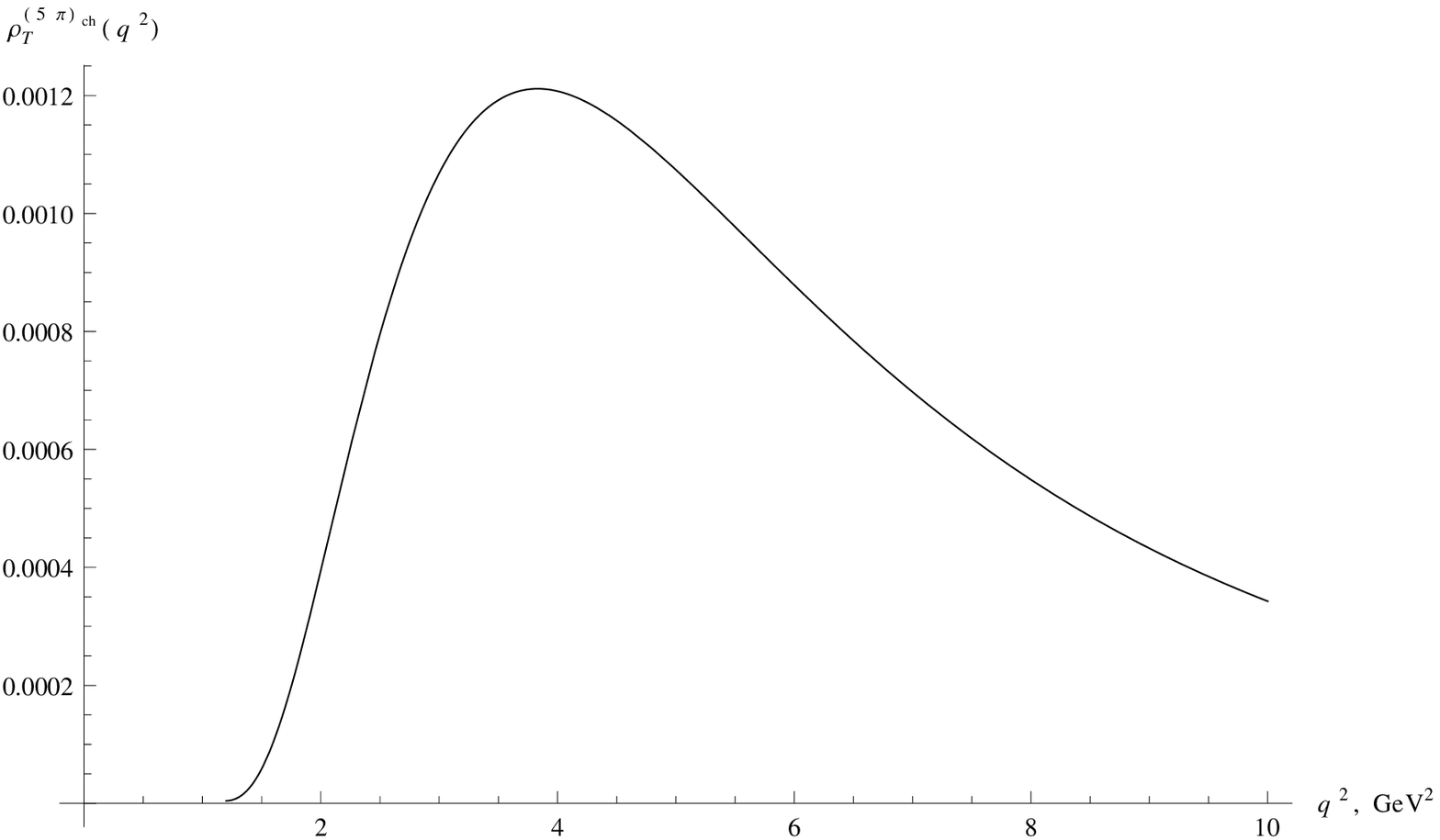} 
\par\end{centering}

\caption{Typical diagram for $W\to(5\pi)_{\ch}$ transition (left figure) and
transverse spectral function $\rho_{T}^{(5\pi)_{\ch}}(q^{2})$ (right
figure)\foreignlanguage{russian}{\label{fig:rhoT5}}}
\end{figure}

If we have five charged $\pi$-mesons in the final state (i.e. $\R=\left(5\pi\right)_{\ch}=\pi^{+}\pi^{+}\pi^{+}\pi^{-}\pi^{-}$)
the amplitude of $W\to(5\pi)_{\ch}$ transition can be obtained from
resonance model \cite{Kuhn:2006nw} (see diagram shown in fig.\ref{fig:rhoT5}).
Due to the partial conservation of vector current the longitudinal
spectral function is equal to zero, while the transverse one can be
approximated by the expression

\begin{eqnarray}
\rho_{T}^{(5\pi)_{ch}}(s) & \approx & 32\left(\frac{s-25m_{\pi}^{2}}{s}\right)^{10}\frac{1-1.65s+0.69s^{2}}{\left[\left(s+2.21\right)^{2}-4.69\right]^{3}},
\end{eqnarray}
 where $s$ is measured in $\mathrm{GeV}^{2}$ (the spectral function
itself is dimensionless). The dependence of this function on $q^{2}$
is shown in fig.\ref{fig:rhoT5}.

\begin{figure}
\includegraphics[width=0.4\textwidth]{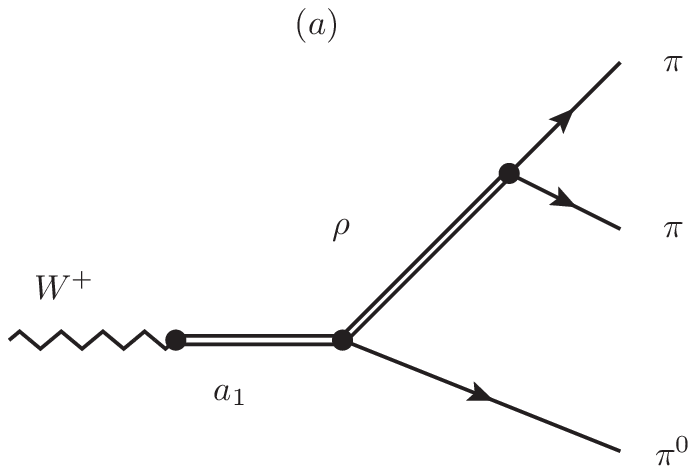} \includegraphics[width=0.4\textwidth]{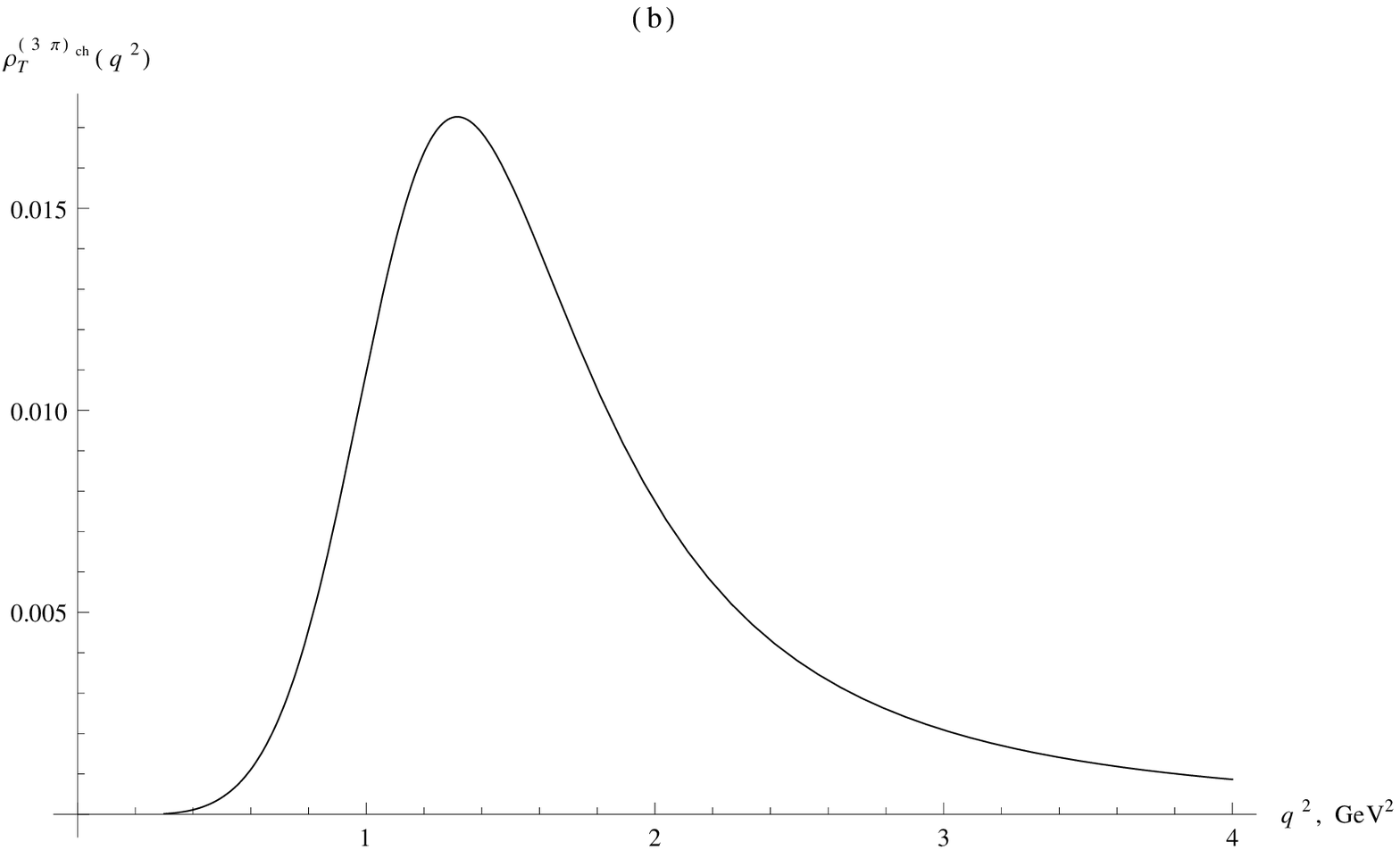}

\caption{Typical diagram for $W\to(3\pi)_{\ch}$ transition (left figure) and
transverse spectral function $\rho_{T}^{(3\pi)_{\ch}}$ (right figure)\label{fig:rhoT3}}
\end{figure}

It should be noted that in addition to direct decays $J/\psi$ and
five charged $\pi$-mesons can be produced in the reaction $B_{c}\to\psi(2S)+(3\pi)_{\ch}\to J/\psi+(5\pi)_{\ch}$,
so the spectral function of $(3\pi)_{\ch}$ state should also be considered.
As in the previous case, this transition can be described in the framework
of resonance model with diagrams shown in fig.\ref{fig:rhoT3}(a)
(see ref.\cite{Kuhn:1990ad}). The contribution of longitudinal spectral
function can be neglected, and the transverse one is approximated
by the expression (see fig.\ref{fig:rhoT3}(b) )

\begin{eqnarray*}
\rho_{T}^{(3\pi)_{ch}}\left(s\right) & \approx & 2.93\times10^{-5}\left(\frac{s-9m_{\pi}^{2}}{s}\right)^{4}\frac{1+190s}{\left[\left(s-1.06\right)^{2}+0.48\right]^{2}}.
\end{eqnarray*}
 Experimentally these two processes can be easily separated.

Formfactors of $B_{c}$-meson transition into vector or pseudoscalar
quarkonium can be calculated, for example, in the framework of QCD
sum rules or various potential models. In our work we use formfactor
sets presented in papers \cite{Kiselev:2002vz} (denoted hereafter
as SR), \cite{Kiselev:2002vz,Kiselev:1993eb,Kiselev:1994ay} (PM1)
and \cite{Ebert:2003cn} (PM2). The values of these formfactors at
points $q^{2}=0$ and $q^{2}=q_{\max}^{2}=(M_{1}^{2}-M_{2}^{2})/(2M_{1})$
are given in tables \ref{tab:ffV} and \ref{tab:ffP}. It can be easily
seen from these tables that for ground quarkonium states formfactors'
behaviour for different models is similar, while predictions for excited
$\psi(2S)$-meson of PM2 model differ dramatically from others. For
example, PM2 model formfactors decrease with the increase of the squared
transferred momentum, while formfactors from SR and PM1 models increase.
Such difference can be probably explained in the following way: in
the framework of potential models $B_{c}$-meson formfactors are determined
from the overlap of initial and final quarkonia wave functions; the
wave function of $\psi(2S)$-meson (in contrast to ground states $J/\psi$
and $B_{s}^{(*)}$) has a node that leads to such unusual behaviour
of the formfactors. It will be shown later that such effect leads
to large difference between different models' predictions for $q^{2}$-distributions
in $B_{c}\to\psi(2S)+(5\pi)_{\ch}$ decays.

\begin{table}
\begin{centering}
\begin{tabular}{|c|c|c|c|c|c|c|c|}
\hline 
\multirow{2}{*}{mode} & \multirow{2}{*}{$F$} & \multicolumn{2}{c|}{SR} & \multicolumn{2}{c|}{PM1} & \multicolumn{2}{c|}{PM2}\tabularnewline
\cline{3-8} 
 &  & $q^{2}=0$  & $q^{2}=q_{\max}^{2}$  & $q^{2}=0$  & $q^{2}=q_{\max}^{2}$  & $q^{2}=0$  & $q^{2}=q_{\max}^{2}$\tabularnewline
\hline 
\hline 
 & $V$  & 1  & 2.1  & 0.94  & 1.9  & 0.49  & 1.3\tabularnewline
\hline 
$B_{c}\to J/\psi\R$  & $A_{0}$  & 0.6  & 1.6  & 0.66  & 1.8  & 0.42  & 1.\tabularnewline
\hline 
 & $A_{1}$  & 0.63  & 1.3  & 0.66  & 1.3  & 0.5  & 0.87\tabularnewline
\hline 
 & $A_{2}$  & 0.69  & 1.4  & 0.66  & 1.3  & 0.73  & 1.3\tabularnewline
\hline 
\hline 
 & $V$  & 0.3  & 0.44  & 0.27  & 0.4  & 0.24  & -0.33\tabularnewline
\hline 
$B_{c}\to\psi(2S)\R$  & $A_{0}$  & 0.15  & 0.28  & 0.16  & 0.31  & 0.24  & 0.011\tabularnewline
\hline 
 & $A_{1}$  & 0.14  & 0.21  & 0.15  & 0.22  & 0.17  & -0.01\tabularnewline
\hline 
 & $A_{2}$  & 0.13  & 0.19  & 0.12  & 0.18  & 0.14  & 0.64\tabularnewline
\hline 
\hline 
 & $V$  & 13  & 17  & 13  & 17  & ---  & ---\tabularnewline
\hline 
$B_{c}\to B_{s}^{*}\R$  & $A_{0}$  & 0.93  & 1.4  & 1.  & 1.5  & ---  & ---\tabularnewline
\hline 
 & $A_{1}$  & 0.69  & 0.09  & 0.71  & 0.92  & ---  & ---\tabularnewline
\hline 
 & $A_{2}$  & -2.3  & -3.  & -3.5  & -4.6  & ---  & ---\tabularnewline
\hline 
\end{tabular}
\par\end{centering}

\caption{Formfactors of $B_{c}\to V$ transition\label{tab:ffV}}
\end{table}

\begin{table}
\begin{centering}
\begin{tabular}{|c|c|c|c|c|}
\hline 
\multirow{2}{*}{$F$} & \multicolumn{2}{c|}{SR} & \multicolumn{2}{c|}{PM1}\tabularnewline
\cline{2-5} 
 & $q^{2}=0$  & $q^{2}=q_{\max}^{2}$  & $q^{2}=0$  & $q^{2}=q_{\max}^{2}$\tabularnewline
\hline 
\hline 
$F_{0}$  & 1.3  & 1.1  & 1.1  & 0.86\tabularnewline
\hline 
$F_{+}$  & 1.3  & 1.7  & 1.1  & 1.5\tabularnewline
\hline 
\end{tabular}
\par\end{centering}

\caption{Formfactors of $B_{c}\to B_{s}$ transition\label{tab:ffP}}
\end{table}

Numerical values of $a_{1}$ coefficient for $B_{c}\to\psi^{(')}+\R$
and $B_{c}\to B_{s}^{(*)}+\R$ decays are 
\begin{eqnarray}
a_{1}\left(m_{c}\right) & = & 1.14
\end{eqnarray}
 and 
\begin{eqnarray}
a_{1}\left(m_{b}\right) & = & 1.2
\end{eqnarray}
 respectively.

Substituting these numbers into relations (\ref{eq:dGV}) and (\ref{eq:dGP})
it is easy to obtain numerical values of the branching fractions of
the decays considered in our article (see table.\ref{tab:Br}). In
order to compare them with the experimental results it is also useful
to consider the ratio of $B_{c}\to V(P)+5\pi$ and $B_{c}\to V(P)+\pi$
branching fractions, where the dependence on the choice of a formfactor
model is partially canceled. In the case of $B_{c}\to J/\psi+\R$
decay these ratios for different models are equal to

\begin{eqnarray*}
\frac{\Br\left[B_{c}\to J/\psi+\left(3\pi\right)_{\ch}\right]}{\Br\left[B_{c}\to J/\psi\pi\right]} & = & \left(\ratioJpppSR\right)_{\mathrm{SR}},\quad\left(\ratioJpppPM\right)_{\mathrm{PM1}},\quad\left(\ratioJpppPMM\right)_{\mathrm{PM2}},
\end{eqnarray*}
 that agrees well with experimental result \cite{LHCb:2012ag} 
\begin{eqnarray}
\frac{\Br_{exp}\left[B_{c}\to J/\psi+(3\pi)_{ch}\right]}{\Br_{exp}\left[B_{c}\to J/\psi\pi\right]} & = & 2.4\pm0.3\pm0.3.
\end{eqnarray}
 and 
\begin{eqnarray}
\frac{\Br\left[B_{c}\to J/\psi+\left(5\pi\right)_{\ch}\right]}{\Br\left[B_{c}\to J/\psi\pi\right]} & = & \left(\ratioJpppppSR\right)_{\mathrm{SR}},\quad\left(\ratioJpppppPM\right)_{\mathrm{PM1}},\quad\left(\ratioJpppppPMM\right)_{\mathrm{PM2}}.
\end{eqnarray}
 For other decays we get 
\begin{eqnarray}
\frac{\Br\left[B_{c}\to\psi(2S)+\left(5\pi\right)_{\ch}\right]}{\Br\left[B_{c}\to\psi(2S)\pi\right]} & = & \left(\ratioJJpppppSR\right)_{\mathrm{SR}},\quad\left(\ratioJJpppppPM\right)_{\mathrm{PM1}},\quad\left(\ratioJJpppppPMM\right)_{\mathrm{PM2}},\label{eq:ratioJJ}\\
\frac{\Br\left[B_{c}\to B_{s}^{(*)}+\left(5\pi\right)_{\ch}\right]}{\Br\left[B_{c}\to B_{s}^{(*)}\pi\right]} & = & \left(\ratioBSALLpppppSR\right)_{\mathrm{SR}},\quad\left(\ratioBSALLpppppPM\right)_{\mathrm{PM1}},
\end{eqnarray}
 where in the last case we take into account contributions from both
vector and pseudoscalar $B_{s}$-mesons. One can notice some interesting
properties of the ratios presented above. First of all, branching
fractions of charmonium production in association with one and five
$\pi$-mesons are comparable with each other, while for $B_{s}$-meson
in the final state the production of large number of $\pi$-mesons
is strongly suppressed. This effect was observed also for $4\pi$
final state (see ref.\cite{Likhoded:2010jr}) and is caused by the
small difference between $B_{c}$- and $B_{s}^{(*)}$-meson masses.
In addition, one can see that PM2 prediction for $B_{c}\to\psi(2S)\R$
decay is about an order of magnitude smaller, than SR and PM1 results.
This difference can be explained by the behaviour of PM2 model formfactors
for excited charmonium state mentioned above.

\begin{table}
\begin{tabular}{|c|c|c|c|}
\hline 
mode  & SR  & PM1  & PM2\tabularnewline
\hline 
\hline 
$B_{c}\to J/\psi\pi$  & $\brJpiSR$  & $\brJpiPM$  & $\brJpiPMM$\tabularnewline
$B_{c}\to J/\psi+(3\pi)_{\ch}$  & $\brJpppSR$  & $\brJpppPM$  & $\brJpppPMM$\tabularnewline
$B_{c}\to J/\psi+\left(5\pi\right)_{\ch}$  & $\brJpppppSR$  & $\brJpppppPM$  & $\brJpppppPMM$\tabularnewline
\hline 
$B_{c}\to\psi(2S)\pi$  & $\brJJpiSR$  & $\brJJpiPM$  & $\brJJpiPMM$\tabularnewline
$B_{c}\to\psi(2S)+\left(5\pi\right)_{\ch}$  & $\brJJpppppSR$  & $\brJJpppppPM$  & $\brJJpppppPMM$\tabularnewline
\hline 
$B_{c}\to B_{s}\pi$  & $\brBSpiSR$  & $\brBSpiPM$  & ---\tabularnewline
$B_{c}\to B_{s}+\left(5\pi\right)_{\ch}$  & $\brBSpppppSR$  & $\brBSpppppPM$  & ---\tabularnewline
\hline 
$B_{c}\to B_{s}^{*}\pi$  & $\brBSSpiSR$  & $\brBSSpiPM$  & ---\tabularnewline
$B_{c}\to B_{s}^{*}+\left(5\pi\right)_{\ch}$  & $\brBSSpppppSR$  & $\brBSSpppppPM$  & ---\tabularnewline
\hline 
\end{tabular}

\caption{Branching fractions of $B_{c}\to V(P)+(5\pi)_{\ch}$ decays\label{tab:Br}}
\end{table}

\begin{figure}
\begin{centering}
\includegraphics[width=0.9\textwidth]{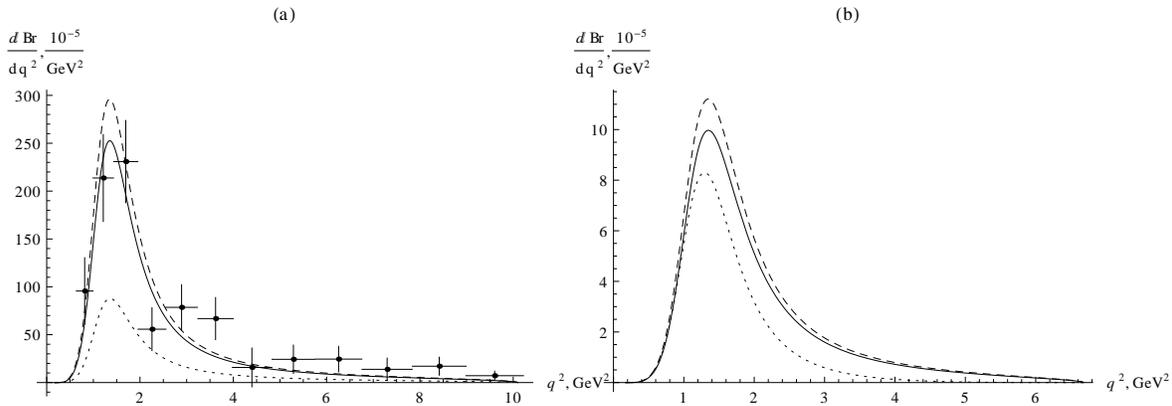} 
\par\end{centering}

\caption{Tranferred momentum distributions for $B_{c}\to J/\psi+(3\pi)_{\ch}$and
$B_{c}\to\psi(2S)+(3\pi)_{\ch}$ decays (left and right figures respectively).
Solid, dashed and dotted lines correspond to formfactor sets SR, PM1,
and PM2 \label{fig:J3pi} }
\end{figure}

Using relations (\ref{eq:dGV}), (\ref{eq:dGP}) one can also obtain
$q^{2}$-distributions for the branching fractions considered in our
article. In the case of $B_{c}\to J/\psi+(3\pi)_{\ch}$ and $B_{c}\to\psi(2S)+(3\pi)_{\ch}$
decays these distributions are shown in fig.\ref{fig:J3pi}. Available
experimental results from \cite{LHCb:2012ag} are shown in the left
plot of this figure with dots. One can see that theoretical predictions
based on SR formfactor model are in good agreement with the experimental
data.

\begin{figure}
\centering{}\includegraphics[width=0.9\textwidth]{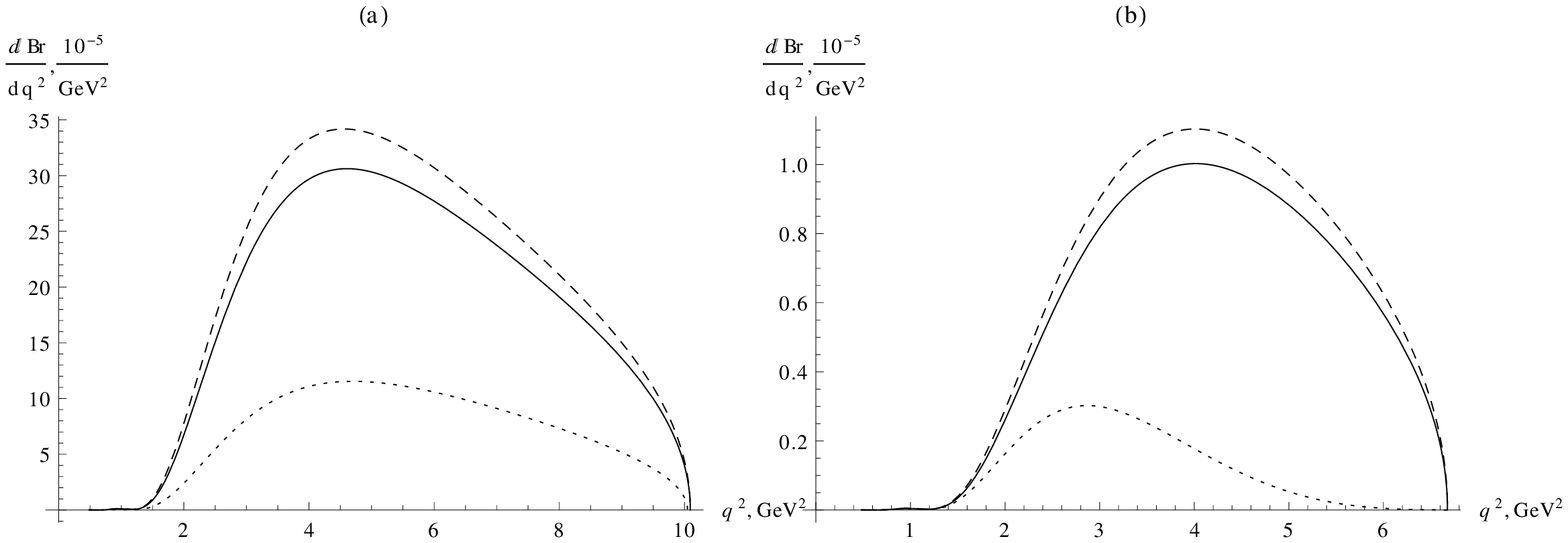} \caption{Transferred momentum distributions of $B_{c}\to J/\psi+(5\pi)_{\ch}$
(left figure) and $B_{c}\to\psi(2S)+(5\pi)_{ch}$ (right figure).
Notations are same as in fig.\ref{fig:J3pi}\label{fig:J5pi} }
\end{figure}

\begin{figure}
\centering{}\includegraphics[width=0.9\textwidth]{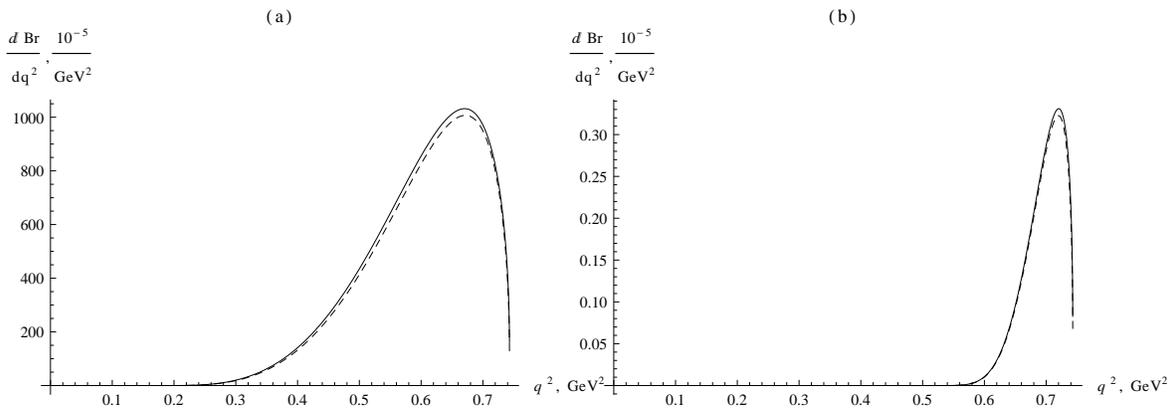} \caption{Trasferred momentum distributions for $B_{c}\to B_{s}^{(*)}+(3\pi)_{\ch}$
(left figure) and $B_{c}\to B_{s}^{(*)}+(5\pi)_{\ch}$ (right figure).
Solid and dashed curves correspond to fromfactor sets SR and PM1\label{fig:BSALL} }
\end{figure}

Similar distributions for $B_{c}\to\psi^{(')}+(5\pi)_{\ch}$ and $B_{c}\to B_{s}^{(*)}+(n\pi)_{\ch}$
decays are shown in figs.\ref{fig:J5pi} and \ref{fig:BSALL} respectively.
One can easily see that for $B_{c}\to J/\psi+(5\pi)_{\ch}$ the results
of different formfactor models agree with each other up to overall
normalization. In the case of $B_{c}\to\psi(2S)+(5\pi)_{\ch}$ decay,
on the other hand, PM2 model prediction differs strongly from SR and
PM1 results. This difference is caused by the behaviour of the formfactors
mentioned above.

\section{Conclusion}

$B_{c}$-meson, i.e. heavy quarkonium build from $b$- and $c$-quarks,
was discovered by ALEPH collaboration in 1997 \cite{Barate:1997kk}.
Reported values of its mass and lifetime are in excellent agreement
with theoretical predictions \cite{Gershtein:1994jw}. The situation
with branching fractions of different decays (e.g $B_{c}\to J/\psi\ell\nu$
or $B_{c}\to J/\psi\pi$) is much worth. It is clear that additional
investigation of this question is required.

In our paper $B_{c}$-meson decays into heavy quarkonium ($J/\psi$,
$\psi(2S)$ or $B_{s}^{(*)}$) and a set of charged $\pi$-mesons
$\left(n\pi\right)_{\ch}$ are considered. In the framework of QCD
factorization theorem the amplitude of these processes can be splitted
into two independent parts that describe $B_{c}\to V(P)W$ vertex
and $W^{*}\to n\pi$ transition. The first part is expressed through
$B_{c}$-meson formfactors, that can be calculated using different
nonperturbative methods: QCD sum rules, various potential models,
etc. Information about $W^{*}\to5\pi$ transition amplitude, on the
other hand, can be obtained from theoretical and experimental analysis
of $\tau$-lepton decays $\tau\to\nu_{\tau}+n\pi$. Using this procedure
we calculated branching fractions and distributions over invariant
mass $q^{2}=M_{(n\pi)}^{2}$ for $n=3$ and $5$. It is shown that
in the case of $B_{c}\to J/\psi+(3\pi)_{\ch}$decay our predictions
agree well with experimental results, presented in \cite{LHCb:2012ag}.
Other considered in our article decays were not observed yet, but
their experimental investigation can be expected in the nearest future.

It should be also noted, that experimental analysis of considered
in our article decays could be useful for studying hadronization of
virtual $W$-boson into a set of light mesons. In the ratio 
\begin{eqnarray}
\rho_{T}(q^{2})\sim(d\Br\left[B_{c}\to V(P)+n\pi\right]/dq^{2})/(d\Br\left[B_{c}\to V(P)+\ell\nu\right]/dq^{2})
\end{eqnarray}
all dependence on $B_{c}$-meson formfactors is canceled, so it describes
only $W\to n\pi$ transition. Previously this function was studied
only in $\tau$-lepton decays, but in $B_{c}$-meson decays one can
probe higher values of transfered momentum.

The author would like to thank A.K. Likhoded, T. Skwarnicki and D.V.
Filippova for useful discussions and help with preparation of this
article. The work was financially supported by Russian Foundation
for Basic Research (grant \#10-00061a) and the grant of the president
of Russian Federation (grant \#MK-3513.2012.2).

\end{document}